\documentclass[aps,preprint]{revtex4}%
\usepackage{amsfonts}
\usepackage{amsmath}
\usepackage{amssymb}
\usepackage{graphicx}%
\providecommand{\U}[1]{\protect\rule{.1in}{.1in}}

\begin{document}
\preprint{gr-qc}
\title[Short title for running header]{A non-singular universe with vacuum energy}
\author{M.B. Altaie}
\affiliation{Department of Physics, Yarmouk University, 21163 Irbid, Jordan}
\author{Usama Al-Ahmad}
\affiliation{Model School, Yarmouk University, 21163 Irbid, Jordan}
\pacs{PACS number}

\begin{abstract}
We construct a model for the universe based on the existence of quantum fields
at finite temperature in the background of Robertson-Walker spacetime in
presence of a non-zero cosmological constant. We discuss the vacuum regime in
the light of the results obtained through previous studies of the
back-reaction of massless quantum fields in the static Einstein universe, and
we argue that an adiabatic vacuum state and thermal equilibrium is achieved
throughout this regime. Results shows that such a model can explain many
features of the early universe as well as the present universe. The model is
free from the basic problems of the standard Friedmann cosmology, and is
non-singular but involves a continuous creation of energy at a rate
proportional to the size of the universe, which is lower than that suggested
by the steady-state cosmology.

\end{abstract}
\eid{identifier}
\date[Date text]{date}
\startpage{1}
\endpage{102}
\maketitle
\tableofcontents

\section{Introduction}

The Friedmann models for the universe which were deduced from the Einstein
field equations with a vanishing cosmological constant and vanishing covariant
derivative of the energy-momentum tensor, described an expanding universe that
starts from a singularity at $t=0.$This develops into a universe that is
either: (a) ever expanding with acceleration if the curvature is negative
($k=-1$), or (b) ever expanding with an ultimate constant speed if the
geometry of the spatial section is flat ($k=0$), or (c) expanding until
reaching a maximum radius, then collapsing under its own gravitational field
to an ultimate singularity, a model which is described as a closed spacetime
with positive curvature ($k=1$). These models found their applications in two
main observational achievements of cosmology during of the last century; the
Hubble discovery of the redshift-distance relationship, and the discovery of
the Cosmic Microwave Background radiation (CMB).

The existence of a homogenous and isotropic CMB was a prediction of the Gamow
and collaborators's big bang theory, which has assumed that the universe
started as a homogeneous and isotropic distribution of particles in thermal
equilibrium at very high temperatures. As the universe expanded and cooled
down, massive particles decoupled from thermal equilibrium. Accordingly
radiation was released as electrons combined with hydrogen and helium nuclei
to form atoms. These radiations are believed to have left a relic which is the
detected CMB. This scenario was further developed and refined and was called
the Standard Big Bang model (SBB).

The SBB utilized statistical physics, particle physics and the standard
general relativity in which particles were not considered in their field
theoretic description, but rather were dealt with through their general
phenomenological description. On the other hand the SBB model did not start
with the universe at $t=0$ since physics cease to work in such a limit. For
these reasons some basic questions remained unanswered; the existence of the
initial singularity posed a challenge for the physicists to define the initial
conditions for the universe. Indeed it was only assumed that the universe
originated from vacuum without giving any explanation for such a birth.
Quantum effects were not investigated within the standard model and therefore,
some important contribution is missing from the SBB model.

Recent analysis of the main features of the CMB suggested that the universe is
highly homogeneous and isotropic on very large scale and that, if it is to be
described geometrically by the Friedmann solutions, then it has to be nearly
flat (i.e. $k=0$) \cite{Bennett}. This result was confirmed by subsequent
observations from WMAP by Spergel et al. \cite{Spergel} and also later by
Dunkley et al. \cite{Dunkley}. This conclusion was based on the fact that
direct observational data suggests that the total matter and radiation density
in the universe is about the same as the critical density needed to flatten
the universe. On the other hand, it well known that the SBB model suffers from
some basic shortcomings, these are: the horizon, the flatness, the magnetic
monopoles and the formation of large cosmic structures problems. To resolve
these problems Guth \cite{Guth1} was the first to suggest that the universe
may have experienced a state of inflation during very early times of its
development, such that the flat geometry of the universe was attained at a
very early stage. This theory was further developed into a main trend in
cosmology where a diversity of inflationary models were suggested (for a
comprehensive review see \cite{Liddle}). A model of inflation typically
amounts to choosing a form for the potential, perhaps supplemented with a
mechanism for bringing inflation to an end, and perhaps may involve more than
one scalar field. In an ideal world the potential would be predicted from
fundamental particle physics, but unfortunately there are many proposals for
possible forms. It has become customary to assume that the potential can be
freely chosen, and then one seek to constrain it with observations.

However, some authors \cite{H&W} argue that explanations provided by inflation
for the homogeneity, isotropy, and flatness of our universe are not
satisfactory, and that a proper explanation of these features will require a
much deeper understanding of the initial state of our universe. On the other
hand, and although inflationary models are spectacularly successful in
providing an explanation of the deviations from homogeneity, these authors
point out that the fundamental mechanism responsible for providing deviations
from homogeneity, namely, the evolutionary behavior of quantum modes with
wavelength larger than the Hubble radius, will operate whether or not
inflation itself occurs. However, if inflation did not occur, one must
directly confront the issue of the initial state of modes whose wavelength was
larger than the Hubble radius at the time at which they were born. Under some
simple hypotheses concerning the \textquotedblright birth
time\textquotedblright\ and initial state of these modes (but without any fine
tuning), it is shown that non-inflationary fluid models, in the extremely
early universe, would result in the same density perturbation spectrum and
amplitude as in inflationary models, although there would be no
\textquotedblright slow roll\textquotedblright\ enhancement of the scalar
modes\cite{H&W}. Other authors believe that inflationary theories are
incomplete since it does not deal with basic puzzles such as the initial
singularity, nor with dark energy indicated by the recent observations
\cite{Turok}.

Ozer and Taha \cite{O&T} devised a model universe free from the basic problems
of the SBB model by assuming a universe predominantly kept at the state of
critical density and, accordingly the scale factor of their model is given by%

\begin{equation}
R^{2}(t)=R_{0}^{2}+t^{2}. \label{q1}%
\end{equation}

On the other hand Chen-Wu \cite{C&W} adopted the prescription that the
cosmological constant varies like $1/a^{2}(t).$Their study resulted in a model
that has the same form for the variation of the scale factor as in the SBB
model, therefore would have no problem integrating the Gamow explanation of
natural abundance, but it would again lead to a singular universe this time
with continuous particle creation at a rate comparable to that suggested by
the steady state theory. The Chen-Wu suggestion was shown to explain the
cosmological constant problem through a phenomenological approach. The good
feature of the Ozer and Taha model is the fluent removal of the SBB problems
without the need for an inflation stage. However it remains that the
assumption of having a universe starting up with a density exactly equal to
the critical density will surely need justification.

It is quite possible that the scale factor do not follow a monotonic behavior
during the whole history of the universe, rather it is quite expectable that
it follows different schemes during the different stages of the development of
the universe. It is certain now that the universe has passed through many
phase-transition states, that had different variations of the scale factor. If
so, then one can say that the Friedmann models will stand as a simplified
picture for the development of the universe and that it is good for describing
the basic theme only. The generally accepted scenario now stipulates that
there are at least three regimes: the vacuum-dominated regime which can be
described by the control of the cosmological constant, the radiation-dominated
regime during which massive particles were in an ultra-relativistic state in
thermal equilibrium with radiation, and the matter-dominated regime which
developed when particles attained non-relativistic states and settled to form
a dust-like fluid.

In this paper we construct a model for the universe with the background
geometry of Robertson-Walker metric endowed with quantum fields at finite
temperature. We assume the presence of a non-zero cosmological constant and
seek the solution of the Einstein field equation. Therefore, the present model
does not satisfy the standard Friedmann paradigm, and in order to construct
the model we utilize the results of previous studies and calculations of the
vacuum energy density in the static Einstein universe. The universe is shown
to have a violent start from a non-singular Plank sized patch developing
through the interaction of vacuum energy and curvature into a very hot spot
and then transiting smoothly into a thermal universe that coasts into the
present one. The model is free from the standard problems of the SBB based on
the standard closed Friedmann model including the initial singularity which
get smoothed-out by the quantum-vacuum effects.

\section{The static Einstein universe}

The investigation of the back-reaction effect of massless quantum fields at
finite temperatures in the background of the static Einstein universe resulted
in a relationship between the radius of the Einstein universe and its
temperature showing some interesting features \cite{Altaie1}. First, all
solutions were shown to posess two regimes, the vacuum (Casimir) regime,
through which the temperature rises from zero sharply reaching a maximum value
of order of 10$^{32}K$ within a very small change of radii, and a Planck
regime, through which the temperature decays exponentially to zero following a
Planckian behavior (see Fig. 1). Second, it was shown that the universe has no
singular state; no static Einstein universe can be singular, rather the ground
state is at zero temperature with a non-zero radius of order of Planck length.
Third, a background (Tolman) temperature was calculated for states of Einstein
universe with large radii, and it was found that the background temperature
equals the observed value of $2.73$ K at a radius nearly two order of
magnitudes as compared to the present Hubble length.

It is well known the static Einstein universe is a solution of the field
equations with a cosmological constant, that can always be adjusted to balance
the gravitational attraction of matter contained within the spatial section of
the universe. The study of the variation of the value of the cosmological
constant with temperature for successive states of the Einstein universe
resulted in showing that the cosmological constant has large and nearly
constant value during the Casimir regime, decaying according to the inverse
square law ($\Lambda\sim1/a^{2}$) during the Planck regime (see Fig. 2). The
exceptional case was with the minimally coupled massless scalar field, where
the cosmological constant was shown to have an infinite value at start,
decaying exponentially to small values already within the Casimir regime
\cite{Altaie2}.

\section{A time-dependent cosmological constant}

The above mentioned results are quite motivating to consider a more realistic
case of a time-dependent model in which the cosmological constant is changing
with the scale factor according to the inverse square law.

In order to include any and all sources of energy that would contribute to the
total energy of the universe, let us write the Friedmann equations in the
following form%

\begin{equation}
\left(  \frac{\overset{.}{a}}{a}\right)  ^{2}+\frac{k}{a^{2}}=\frac{8\pi}%
{3}\left(  \rho_{\text{m, r}}\right)  _{^{\text{eff}}}+\frac{\lambda
_{^{\text{eff}}}}{3}, \label{q2}%
\end{equation}
where $(\rho_{\text{m.r}})_{\text{eff}}$ is the energy density for mater and
radiation, and $\lambda_{^{\text{eff}}}$ represents any and all contributions
coming from the cosmological constant or any other source of energy-momentum
density including the vacuum energy density. In other form the Friedmann
equation can be written as%

\begin{equation}
\left(  \frac{\overset{.}{a}}{a}\right)  ^{2}+\frac{k}{a^{2}}=\frac{8\pi}%
{3}\left(  T_{0}^{0}\right)  _{\text{total}}+\frac{\lambda}{3}, \label{q3}%
\end{equation}
where $\left(  T_{0}^{0}\right)  _{\text{total}}=\rho_{\text{m, r}}%
+\rho_{\text{v}},$with $\rho_{\text{v}}=\left\langle 0|T_{0}^{0}%
|0\right\rangle _{tot}$ is the total vacuum energy density. Eq. (\ref{q3}) can
be re-written as%

\begin{equation}
6\left[  \left(  \frac{\overset{.}{a}}{a}\right)  ^{2}+\frac{k}{a^{2}}%
-\frac{8}{3}\pi\left(  2\rho_{\text{m, r}}-\frac{3}{8\pi}\left(
\frac{\overset{.}{a}}{a}\right)  ^{2}\right)  -\frac{2\lambda}{3}\right]
=32\pi\left\langle 0|T_{00}|0\right\rangle _{tot}-\frac{6k}{a^{2}}. \label{q4}%
\end{equation}

Now if%

\begin{equation}
\left[  2\rho_{\text{m, r}}-\frac{3}{8\pi}\left(  \frac{\overset{.}{a}}%
{a}\right)  ^{2}\right]  =\left(  \rho_{\text{m, r}}\right)  _{\text{eff}%
}=\rho_{c}\text{ } \label{q5a}%
\end{equation}
and%

\begin{equation}
2\lambda=\lambda_{^{\text{eff}}}, \label{q5b}%
\end{equation}
then we obtain%

\begin{equation}
6\left[  \left(  \frac{\overset{.}{a}}{a}\right)  ^{2}+\frac{k}{a^{2}}%
-\frac{8}{3}\pi\left(  \rho_{\text{m, r}}\right)  ^{\text{eff}}-\frac
{\lambda^{\text{eff}}}{3}\right]  =32\pi\left\langle T_{00}\right\rangle
_{tot}-\frac{6k}{a^{2}}. \label{q6}%
\end{equation}
Comparing the left hand side of (\ref{q6}) with (\ref{q2}) we get%

\begin{equation}
\frac{6k}{a^{2}}=32\pi\left\langle 0|T_{0}^{0}|0\right\rangle _{tot}.
\label{q7}%
\end{equation}
for $k=1$ this will give the same result as that obtained in the case of
closed static universe \cite{Altaie1}.

From the basic Einstein field equation and for the Robertson-Walker metric we
can deduce that%

\begin{equation}
T_{0\text{ };\text{ }\mu}^{\mu}=\overset{.}{\rho}+3\left(  \frac{\overset
{.}{a}}{a}\right)  \left(  \rho+p\right)  . \label{q8}%
\end{equation}

Applying the covariant derivative to the Einstein field equation we obtain%

\begin{equation}
\overset{.}{\rho}+3\left(  \frac{\overset{.}{a}}{a}\right)  \left(
\rho+p\right)  =-\frac{1}{8\pi}\overset{.}{\lambda}. \label{q9}%
\end{equation}

This equation will stand as a replacement for the equation of state used by
the standard Friedmann models where the right-hand side is taken to vanish on
the assumption that $\lambda$ is constant.

The second law of thermodynamics requires that%

\[
dE+pdV=TdS,
\]
which means that%

\begin{equation}
\frac{dE}{dt}+p\frac{dV}{dt}=T\frac{dS}{dt}, \label{q10}%
\end{equation}
where $E=m_{0}c^{2}=\left(  \rho V\right)  ,$ \ and $V=2\pi^{2}a^{3}$ is the
volume of the closed universe$.$Therefore%

\begin{equation}
\frac{dE}{dt}=6\pi^{2}a^{2}\rho\frac{da}{dt}+2\pi^{2}a^{3}\frac{d\rho}{dt},
\label{q11}%
\end{equation}
so that for Eq. (\ref{q10}) we obtain%

\begin{equation}
V\left[  \overset{.}{\rho}+3\left(  \frac{\overset{.}{a}}{a}\right)  \left(
\rho+p\right)  \right]  =T\frac{dS}{dt}, \label{q12}%
\end{equation}
where the dot denotes differentiation with respect to time. Substituting
(\ref{q12}) in (\ref{q9}) we get%

\begin{equation}
TdS=-\frac{V}{8\pi}d\lambda. \label{q13}%
\end{equation}

Therefore, a variable $\lambda$ may solve the entropy problem without the
introduction of specific fields or irreversible processes. The idea that
$\lambda$ may be variable has previously been suggested by several authors
\cite{O&T}, \cite{Altaie2} and \cite{Abd}.

For the homogeneous isotropic model with Robertson-Walker metric with
$\rho=\rho\left(  t\right)  ,$ $p=p\left(  t\right)  $ and $\lambda
=\lambda\left(  t\right)  ,$ Eq. (\ref{q2}), (\ref{q5a}) and (\ref{q5b}) yield%

\begin{equation}
\left(  \frac{\overset{.}{a}}{a}\right)  ^{2}=\alpha^{-1}\left(
\rho_{\text{m, r}}\left(  t\right)  +\Lambda\left(  t\right)  \right)
-\frac{k}{2a^{2}}, \label{q14}%
\end{equation}
where $\alpha=\left(  3/8\pi\right)  $ and $\Lambda=\left(  \lambda
/8\pi\right)  .$

Thus $dS/dt\geq$ $0$ requires that $d\Lambda/dt\leq0$ for all $t\geq0.$ Then
From Eq.(\ref{q14}) we may interpret $\Lambda\left(  t\right)  $ as a vacuum
or cosmological energy density.\bigskip

Now from (\ref{q10}) and (\ref{q13}) we get%

\[
\frac{d\left(  \rho V\right)  }{dt}+p\frac{dV}{dt}+\frac{V}{8\pi}%
\frac{d\lambda}{dt}=0,
\]
which means that%

\begin{equation}
\frac{d\left(  \rho a^{3}\right)  }{dt}+p\frac{da^{3}}{dt}+a^{3}\frac
{d\Lambda}{dt}=0. \label{q15}%
\end{equation}

Taking $p=\frac{1}{3}\rho$ for radiation filled universe we get%

\begin{equation}
\frac{d\rho}{da}+\frac{d\Lambda}{da}+\frac{4\rho}{a}=0. \label{q16}%
\end{equation}

To solve this equation we consider the general form%

\begin{equation}
\frac{dy}{dx}+P\left(  x\right)  y=Q\left(  x\right)  , \label{q17}%
\end{equation}
where the solution is given by%

\begin{equation}
y=\exp\left[  -I\left(  x\right)  \right]  \int Q\left(  x\right)  \exp\left[
I\left(  x\right)  \right]  dx+b\exp\left[  -I\left(  x\right)  \right]  ,
\label{q18}%
\end{equation}
where%

\begin{equation}
I\left(  x\right)  =\int P\left(  x\right)  dx, \label{q19}%
\end{equation}
and $b$ is a constant. Here we have $P\left(  a\right)  =\frac{4}{a}$ so that%

\begin{equation}
I\left(  a\right)  =\int_{a_{0}}^{a}P\left(  a\right)  da==\ln\left(  \frac
{a}{a_{0}}\right)  ^{4} \label{q20}%
\end{equation}
Also%

\begin{equation}
Q\left(  a\right)  =-\frac{d\Lambda}{da}. \label{q21}%
\end{equation}
consequently we obtain the solution%

\begin{equation}
\rho=\rho_{0}\left(  \frac{a_{0}}{a}\right)  ^{4}-\left(  \frac{1}{a}\right)
^{4}\int_{a_{0}}^{a}\left(  a%
\acute{}%
\right)  ^{4}\frac{d\Lambda}{da%
\acute{}%
}da%
\acute{}%
. \label{q22}%
\end{equation}
We shall take $\rho_{0}$ and $a_{0}$ to be the values of the energy density
and scale factor at $t=0$.

If the total energy of the universe, including the vacuum energy, is to be
constant then we have to take $dE=0$, which implies that%

\begin{equation}
TdS+a^{3}d\Lambda=0. \label{q23}%
\end{equation}

One observes that the conditions $\overset{.}{\Lambda}\leq0$ and $\overset
{.}{a}\geq0$ imply that the integral in (\ref{q22}) is negative so that the
decrease of $\Lambda$ as $a$ increases generates a positive contribution to
$\rho.$ Thus the cosmological energy density $\Lambda$ is continually depleted
and transformed into radiation energy density in accordance with (\ref{q22}).
This consolidates the conjecture of Altaie and Setare \cite{Altaie2}.

Following \cite{O&T} we take the initial time $t=0$ to be the moment when
$\overset{.}{a}=0$. Then if $\rho_{0}=0,$ Eq. (\ref{q22}) is still valid and
the empty curved space-time metric which is governed by%

\begin{equation}
R_{\nu}^{\mu}-\frac{1}{2}g_{\nu}^{\mu}R=-8\pi\left[  T_{\nu}^{\mu}\right]
^{\left(  \text{vac}\right)  }\text{, \ \ \ }t=0, \label{q24}%
\end{equation}
soon will generate a non-empty universe, i.e., one with non-vanishing
radiation density $\rho.$ One might thus take Eq. (\ref{q22}) to imply that
radiation is being created out of the space-time curvature all times. The
physical picture one has is therefore the one in which the universe is
continuously created by the unfolding of space-time curvature. This is
different from the continuous creation in the Bondi-Gold-Hoyle \cite{Bondi}
model which was suggested in the context of steady-state cosmology. The
steady-state model is also based on classical general relativity\textbf{\ }%
with modified $T_{\mu\nu}$. The modification is by the addition of a covariant
term which was chosen so that $\rho,$ $p$ and $H$ \ remain constant throughout
the de Sitter expansion of the universe. Continuous creation and the absence
of an initial singularity are features of steady-state cosmology that are
shared by the present model.

In the present formulation the cosmological energy density $\Lambda$ is
related to entropy by%

\begin{equation}
TdS+a^{3}d\Lambda=0. \label{q25}%
\end{equation}
This equation may in fact be interpreted as an expression of the constancy of
the total entropy of the cosmos, substance and spacetime, i.e.,%

\begin{equation}
dS+dS_{c}=0, \label{q26}%
\end{equation}

where%

\begin{equation}
dS_{c}=\frac{a^{3}}{T}d\Lambda, \label{q27}%
\end{equation}
is the change in the entropy of the curved spacetime. On the long run
($t\rightarrow\infty$) this change flattens the spacetime. Under these
conditions one would intuitively expect a decrease in the entropy
$S_{\text{c}}$ since the number of degrees of freedom one might associate with
a state of high curvature should be larger than those associated with an
almost flat space.

We note that in a universe of pure radiation the choice of $\Lambda$
completely determines the model, since $\Lambda$ determines $\rho$ by Eq.
(\ref{q22})$,$ leaving Eq. (\ref{q14}) to be solved for $a\left(  t\right)  .$
However when, in addition to radiation matter is present, extra assumptions
are needed to uniquely determine the model. In this case $\rho=\rho_{\text{m}%
}+\rho_{\text{r}},$ where $\rho_{\text{m}}$ is the rest-mass energy density
and $\rho_{\text{r}}$ is the energy density of radiation and relativistic
matter (to be referred simply as radiation in what follows). Then Eq.
(\ref{q15}) reads%

\begin{equation}
d\left(  \rho_{\text{m}}a^{3}\right)  +d\left(  \rho_{\text{r}}a^{3}\right)
+pda^{3}=-a^{3}d\Lambda. \label{q28}%
\end{equation}

A plausible assumption that may readily be made is that the processes
responsible for changes in rest-mass energy density are, except for the matter
creation period, much slower than those responsible for creation of radiation,
(see \cite{O&T}); i.e.,%

\begin{equation}
\left\vert \frac{d}{dt}\left(  \rho_{\text{m}}a^{3}\right)  \right\vert
\ll\left\vert \frac{d}{dt}\left(  \rho_{\text{r}}a^{3}\right)  \right\vert .
\label{q29}%
\end{equation}

We also make the assumption that the pressure of the universe, under these
conditions, is caused by its radiation%

\begin{equation}
p=p_{\text{r}}=\frac{1}{3}\rho_{\text{r}}. \label{q30}%
\end{equation}
Eq. (\ref{q15}) then yield%

\begin{equation}
d\left(  \rho_{\text{r}}a^{3}\right)  +\frac{1}{3}\rho_{\text{r}}da^{3}%
=-a^{3}d\Lambda, \label{q31}%
\end{equation}
and%

\begin{equation}
d\left(  \rho_{\text{m}}a^{3}\right)  \thickapprox0, \label{q32}%
\end{equation}
implying that Eq. (\ref{q15}) may be assumed to be approximately valid for the
radiation component in both the radiation-dominated as well as the
matter-dominated eras. This may alternatively be interpreted in the sense that
a substance is created, by the unfolding of curved spacetime, as massless
radiation. Yet another interpretation could be that the change in entropy of
non-relativistic matter is much less than that of radiation, i.e., $\Delta
S_{\text{m}}\ll\Delta S_{\text{r}},$ since $-a^{3}d\Lambda$ is a measure of
the total change in the entropy.

\section{The model}

We now deduce a particular function for $\Lambda$, thereby defining a specific
cosmological model in the classical class. Our starting point is the
observation that the present value of the energy density of the universe is
close to its critical value $\rho_{c}$. Theoretically the investigations of
the back-reaction effects of quantum fields at finite temperatures indicates
that the density of the universe was always fixed at the critical value%

\begin{equation}
\rho_{c}\left(  t\right)  =\frac{3H^{2}}{8\pi}=\frac{3}{8\pi}\left(
\frac{\overset{\cdot}{a}}{a}\right)  ^{2}. \label{q33}%
\end{equation}

From Eq. (\ref{q14}) we notice that the condition $\rho=\rho_{c}$ requires that%

\begin{align}
\left(  \frac{\overset{.}{a}}{a}\right)  ^{2}  &  =\alpha^{-1}\left(  \rho
_{c}+\Lambda\left(  t\right)  \right)  -\frac{k}{2a^{2}}\nonumber\\
&  =\alpha^{-1}\left(  \alpha\left(  \frac{\overset{.}{a}}{a}\right)
^{2}+\Lambda\left(  t\right)  \right)  -\frac{k}{2a^{2}}. \label{q34}%
\end{align}
So that $\Lambda(t)$ comes to be%

\begin{equation}
\Lambda\left(  t\right)  =\frac{\alpha k}{2a^{2}}. \label{q35}%
\end{equation}

The conditions $\overset{.}{\Lambda}\leq0$ and $\overset{.}{a}\geq0$ then
immediately give $k\geq0$ so that $k=1.$ This is a significant deviation from
the standard model where $\rho=\rho_{c}$ implies $k=0.$ This is an important
feature of the present model, therefore, we can confidently conclude that%

\begin{equation}
\Lambda\left(  t\right)  =\frac{\alpha}{2a^{2}}. \label{q36}%
\end{equation}

The model now is completely specified in its dynamical structure. The rest of
the work will mostly run in similar fashion to that of Ozer and Taha with a
difference by a factor of $2$ in some equations. However, as for the initial
start of the universe we here do not need to follow the assumption of Ozer and
Taha but would rather resort to take the results of previous works which
defined for us the initial radius of the universe as a result of
self-consistency condition applied on the Einstein field equations for a given
quantum field source. The provisions of an equation of state determines the
physical content. In the following sections we present the results of the
calculations with some outlines.

\subsection{The very early universe}

In our model the initial moment $t=0$ has been chosen to coincide with the
state of the universe when its energy content is specified by the presence of
Casimir energy resulting from the high initial curvature. Self-consistency of
the Einstein field equation requires the universe to have a non-zero radius at
$T=0$. This was already calculated in a previous works (see \cite{Altaie3})
There are, initially, no such excitations since $\overset{.}{a}$ $(0)=0$ and
the total energy is locked up in potential form in spacetime curvature. With
$\rho_{0}=0,$ Eqs. (\ref{q22}) and (\ref{q36}) yield%

\begin{equation}
\rho=\frac{\alpha}{2a^{2}}\left(  1-\frac{a_{0}^{2}}{a^{2}}\right)  ,
\label{q40}%
\end{equation}

Note that the condition $\rho_{0}=0$ requires $a_{0}\neq0.$ This implies that
all functions $a,$ $\rho,$ $S,$ $T$ are finite at $t=0$ (as well as for all
finite $t$ as will soon become clear) so that the initial singularity of the
standard model does not exist.

Eq. (\ref{q40}) with $\rho=\rho_{c}=\alpha(\frac{\overset{\cdot}{a}}{a})^{2}$,
may now be solved for $a\left(  t\right)  $ giving%

\begin{equation}
\frac{ada}{\left[  a^{2}-a_{0}^{2}\right]  ^{1/2}}=\left(  \frac{1}{2}\right)
^{1/2}dt. \label{q42}%
\end{equation}
This can be easily integrated to give%

\begin{equation}
a(t)=\left[  a_{0}^{2}+\frac{t^{2}}{2}\right]  ^{1/2}. \label{q43}%
\end{equation}

It is clear that the universe starts accelerating during the Casimir regime
and then soon get to an ultimate speed when it reaches the velocity of light.
In this model we have no inflation but a direct parametric dependence of the
radius on time. Note that in this solution of the field equations
$a\rightarrow\infty$ as $t\rightarrow\infty$ i.e., the model is continuously
expanding although $k=1.$ This is due to the variable cosmological energy
density $\Lambda$ in Eq. (\ref{q34}) which renders the characterization of the
asymptotic behavior being $\Lambda-$dependent. Different choices of $\Lambda,$
for fixed $k$, can give different types of asymptotic behavior. The intrinsic
geometry, for fixed $t,$ is governed by the parameter $k$ and is independent
of $\Lambda.$

The time-dependence of all functions in the model is completely determined by
Eqs. (\ref{q40}) and (\ref{q43}). For the radiation energy density we have%

\begin{equation}
\rho(t)=\frac{\alpha}{2}\left[  \frac{\left(  t/\sqrt{2}\right)  ^{2}}{\left(
a_{0}^{2}+\left(  t/\sqrt{2}\right)  ^{2}\right)  ^{2}}\right]  \label{q45}%
\end{equation}

The radiation temperature $T$ \ is assumed to be related to $\rho$ by%

\begin{equation}
\rho(T)=\frac{1}{30}\pi^{2}N\left(  T\right)  T^{4} \label{q47}%
\end{equation}

The effective number of spin degree of freedom $N\left(  T\right)  $ at
temperature $T$ \ is given by $N\left(  T\right)  =N_{\text{b}}\left(
T\right)  +\frac{7}{8}N_{\text{f}}\left(  T\right)  ,$ where $N_{\text{b}%
}\left(  T\right)  $ and $N_{\text{f}}\left(  T\right)  $ refer to bosons and
fermions respectively. We take $N\left(  T\right)  $ to be constant throughout
the pure radiation era. From Eqs. (\ref{q45}) and (\ref{q47}) we obtain%

\begin{equation}
T(t)=\left(  \frac{15\alpha}{\pi^{2}N}\right)  ^{1/4}\left[  \frac{t^{2}%
/2}{(a_{0}^{2}+t^{2}/2)^{2}}\right]  ^{1/4} \label{q48}%
\end{equation}

This is qualitatively the same result that was obtained by Ozer and Taha. In
terms of the radius $a$ \ the dependence of $T$ \ is given by
\begin{equation}
T(a)=\left(  \frac{15\alpha}{\pi^{2}N}\right)  ^{1/4}\left[  \frac{a^{2}%
-a_{0}^{2}}{a^{4}}\right]  ^{1/4}. \label{q49}%
\end{equation}

Thus, according to this model the universe have a cold start since $T=0$ at
$t=0.$ For small $a_{0},$ this need, however, not to be different from a hot
universe since temperature increases rapidly within a time-scale of order
$a_{0}.$ Fig. 3 shows the qualitative time development of the temperature of
the universe $T$ $\ $according to Eq. (\ref{q49}). Also, from Eqs.
(\ref{q36})\ and (\ref{q43}) we get Fig. 4, which illustrates a qualitative
relationship between $T$\ \ and $\Lambda.$

The maximum temperature $T_{\text{max}}$ is obtained at $t=\sqrt{2}a_{0}$ and
is given by
\begin{equation}
T_{\text{max}}=\left(  \frac{1}{2}\right)  ^{1/4}\left(  \frac{15\alpha}%
{2\pi^{2}Na_{0}^{2}}\right)  ^{1/4},\label{q50}%
\end{equation}

For $t\gg a_{0},$ Eqns. (\ref{q45}) and (\ref{q48}) gives%

\begin{equation}
\rho=\frac{\alpha}{t^{2}} \label{q51}%
\end{equation}
and%

\begin{equation}
T=\left(  \frac{30\alpha}{\pi^{2}N}\right)  ^{1/4}\left[  \frac{1}{t}\right]
^{1/2}. \label{q52}%
\end{equation}

These equations are to be compared to those of the standard model, namely
\begin{equation}
\rho_{\text{SM}}=\frac{\alpha}{\left(  2t\right)  ^{2}}, \label{q53}%
\end{equation}
and%

\begin{equation}
T_{\text{SM}}=\left(  \frac{30\alpha}{\pi^{2}N}\right)  ^{1/4}\left[  \frac
{1}{2t}\right]  ^{1/2}. \label{q54}%
\end{equation}

Thus for $t\geq a_{0}$ the values of the energy density and temperature
attained at a time $t$ \ in the standard model are attained at time $2t$ in
the present model. For $t\thickapprox a_{0},$ Eqs. (\ref{q45}) and (\ref{q48})
coincide with Eqs.(\ref{q53}) and (\ref{q54}) of the standard model. One may,
therefore, conclude that although this model is clearly different from the
standard model in several aspects, such as having cold initiation,
non-adiabaticity and regularity at $t=0,$ it possesses for $T\geq
T_{\text{max}}$ essentially the same thermal history as the standard model.
This is somewhat surprising since the dependence of $a$ on $t$, in the present
model, is completely different from that of the standard model: $a_{\text{SM}%
}\sim t^{1/2}.$ On the other hand we should, however, expect some substantial
deviation in the time-dependence as compared to the standard model if the
cosmological problems of the standard model are to be avoided.

If we have to look at the variation of the cosmological constant with the
temperature of the universe, then we will obtain the temperature dependence
depicted in Fig. 4. The figure shows a dependence which is similar to that we
obtained for the Einstein universe for conformaly coupled scalar field.

In particular, the dependence of the cosmic scale factor $a$ on $t$ determines
the causal structure of the model. The horizon distance $d_{H}\left(
t\right)  $ at time $t$ is the proper distance travelled by light emitted at
$t=0$%

\begin{equation}
d_{H}\left(  t\right)  =a\left(  t\right)  \int_{0}^{t}\frac{dt}{a\left(
t\right)  } \label{q55}%
\end{equation}

For the universe around us to be causally-connected to us at cosmic time $t$
it is necessary that $d_{H}\left(  t\right)  \geq d_{\text{proper}}\left(
t\right)  $ where $d_{\text{proper}}\left(  t\right)  $ is the proper
distance, at time $t$ between our galaxy and another galaxy most distant from
us assuming that our galaxy is at $r=0.$ :%

\begin{equation}
d_{\text{proper}}\left(  t\right)  =a\left(  t\right)  \int_{0}^{r_{\text{max}%
}}\frac{dr}{\sqrt{1-kr^{2}}}=a\left(  t\right)  \left[
\begin{tabular}
[c]{ll}%
$r_{\text{max}},$ & $k=0$\\
$\sinh^{-1}r_{\text{max}},$ & $k=-1$\\
$\sin^{-1}r_{\text{max}}\text{,}$ & $k=1$%
\end{tabular}
\ \ .\right]  \label{q56}%
\end{equation}

For $k=0$ and $k=-1$ the universe is spatially infinite so that $r_{\text{max}%
}=\infty$. This implies that for $k=0$ and $k=-1$ global causal connection,
i.e. causal connection for the whole space, is never established at any finite
time $t.$ The region of the universe which is causally connected at time $t$
is limited in coordinate space to $0\leq r\leq r_{H}\left(  t\right)  ,$ where%

\begin{equation}
\int_{0}^{r_{H}\left(  t\right)  }\frac{dr}{\sqrt{1-kr^{2}}}=\int_{0}^{t}%
\frac{dt%
\acute{}%
}{a\left(  t%
\acute{}%
\right)  }. \label{q57}%
\end{equation}

The horizon (or causality) problem for $k=0$ and $k=-1$ in the standard model
may be formulated only for the currently observed universe, i.e. in non-global terms.

For $k=1$ the universe is spatially finite and $r_{\text{max}}$ $=1.$ It is
then possible to determine the time $t=t_{\text{caus}}$ when the whole
universe is causally connected. This is given by%

\begin{equation}
\int_{0}^{t_{\text{caus}}}\frac{dt%
\acute{}%
}{a\left(  t%
\acute{}%
\right)  }=\int_{0}^{1}\frac{dr}{\sqrt{1-kr^{2}}}=\frac{\pi}{2}. \label{q58}%
\end{equation}

For the standard model with $k=1,$ one finds that by the end of the
radiation-dominated era, $t_{r}=10^{12}$ s, only a small part of the whole
space is causally connected. To see this%

\begin{equation}
\int_{0}^{t_{r}}\frac{dt}{a\left(  t\right)  }=2\left(  \frac{t_{r}}{A^{2}%
}\right)  ^{1/2}. \label{q59}%
\end{equation}

Using $a=At^{1/2}$ in the radiation-dominated era of the standard model where
$A=\left(  2\pi^{2}N/15\alpha\right)  ^{1/4}a_{\text{p}}T_{\text{P}}.$ For
$N=100,$ $a_{\text{p}}=10^{10}$ years and $T_{\text{P}}=2.7$K ,\ the
right-hand side of (\ref{q59}) is $0.014$ so that Eq. (\ref{q57}) with $k=1$
gives $r_{H}\left(  t_{r}\right)  =\sin0.014\thickapprox0.014\ll1.$ Global
causal connection, i.e. up to $r=1$, is realized during the matter-dominated
era as may be verified using \ (\ref{q58})$.$ Thus most of the currently
observable universe was not in causal contact at the end of the
radiation-dominated era. One is then unable to explain the observed isotropy
of the background black-body radiation.

In the present model Eqs. (\ref{q43}) and (\ref{q58}) determine the time
$t_{\text{caus}}$ when global causality is established. One finds%

\begin{equation}
\int_{0}^{t_{\text{caus}}}\frac{dt%
\acute{}%
}{\left(  a_{0}^{2}+\left(  t/\sqrt{2}\right)  ^{2}\right)  ^{1/2}}=\frac{\pi
}{2}, \label{q60}%
\end{equation}

Now let%

\[
\frac{t}{\sqrt{2}}=a_{0}\sinh w\text{ \ \ }\Rightarrow\text{ \ \ \ }%
dt=\sqrt{2}a_{0}\left(  \cosh w\right)  \text{ }dw,
\]
then Eq. (\ref{q60}) becomes%

\begin{equation}
\int_{0}^{w_{\text{caus}}}\frac{\sqrt{2}a_{0}\left(  \cosh w\right)  \text{
}dw}{\left(  a_{0}^{2}+\left(  a_{0}\sinh w\right)  ^{2}\right)  ^{1/2}}%
=\frac{\pi}{2}, \label{q61}%
\end{equation}
which then would yield%

\begin{equation}
t_{\text{caus}}=\sqrt{2}a_{0}\text{ }\sinh\left(  \frac{\pi}{2\sqrt{2}%
}\right)  =1.9a_{0}. \label{q62}%
\end{equation}

Note that for the integral in (\ref{q58}) to converge it is necessary to have
$a_{0}\neq0.$ Eq. (\ref{q62}) indicates that global causal connection in the
present model has been established at a very early time. Thus the present
model does not possess a horizon problem.

We observes that the \textquotedblleft cold era\textquotedblright\ in this
model is restricted to an interval when the causally connected part of the
universe covers a tiny fraction of the whole space. When global causality is
established at $t_{\text{caus}}=$ $1.9a_{0}$ the maximum temperature is
surpassed and the whole universe attains the temperature $T_{\text{caus}}%
.$From Eq. (\ref{q50}), we have:%

\begin{equation}
T_{\text{max}}=\left(  \frac{15\alpha}{4\pi^{2}N}\right)  ^{1/4}\left(
\frac{1}{a_{0}^{2}}\right)  ^{1/4}, \label{q63}%
\end{equation}
also with the help of Eqs. (\ref{q50}) and (\ref{q62}), we get:%

\begin{equation}
T_{\text{caus}}=0.6916\left(  \frac{15\alpha}{\pi^{2}N}\right)  ^{1/4}\left(
\frac{1}{a_{0}^{2}}\right)  ^{1/4}. \label{q64}%
\end{equation}
this shows that%

\begin{equation}
T_{\text{caus}}=0.978T_{\text{max}} \label{q65}%
\end{equation}

\subsection{Radiation and matter}

The very early pure radiation era soon gives way to a period of matter
generation. Throughout this period, $a_{1}\leq a\leq a_{2}$ say, Eq.
(\ref{q28}) is valid. For $a\geq a_{2}$ Eqs. (\ref{q31}) and (\ref{q32}) hold,
so that the total rest-mass energy%

\begin{equation}
E_{\text{m}}=\rho_{\text{m}}a^{3}, \label{q66}%
\end{equation}
remains approximately constant, i.e.,
\begin{equation}
\rho_{\text{m}}=\frac{E_{\text{m}}^{\text{p}}}{a^{3}}. \label{q67}%
\end{equation}

Note that although the proper volume is $2\pi^{2}a^{3}$ we shall, in
accordance with general convention, take it to be simply $a^{3}.$ This is of
no consequence since the measurable quantity is the energy density.

The solution to Eq. (\ref{q31}) for $\rho_{\text{r}}$ with $\Lambda$ given by
Eq. (\ref{q36})\ is%

\begin{equation}
\rho=\rho_{0}\exp\left[  -\ln\left(  \frac{a}{a_{p}}\right)  ^{4}\right]
+\exp\left[  -\ln\left(  \frac{a}{a_{p}}\right)  ^{4}\right]  \int
_{a_{\text{p}}}^{a}\left(  \frac{\alpha}{a%
\acute{}%
^{3}}+\frac{E_{\text{m}}^{\text{p}}}{a%
\acute{}%
^{4}}\right)  \exp\left[  \ln\left(  \frac{a%
\acute{}%
}{a_{p}}\right)  ^{4}\right]  . \label{q68}%
\end{equation}
Now, if $\rho=\rho_{0}$ when $a=a_{0},$ then $\rho$ becomes%

\begin{align}
\rho &  =\left(  \frac{a_{p}}{a}\right)  ^{4}\int_{a_{\text{p}}}^{a}\left(
\frac{\alpha}{a%
\acute{}%
^{\text{ }3}}+\frac{E_{\text{m}}^{\text{p}}}{a%
\acute{}%
^{\text{ }4}}\right)  \left(  \frac{a%
\acute{}%
}{a_{p}}\right)  ^{4}da%
\acute{}%
\nonumber\\
&  =\frac{\alpha}{2a^{2}}\left[  1+\left(  -1-\frac{2E_{\text{m}}^{\text{p}}%
}{\alpha a_{\text{p}}}\right)  \left(  \frac{a_{\text{p}}^{2}}{a^{2}}\right)
\right]  +\frac{E_{\text{m}}^{\text{p}}}{a^{3}}\nonumber\\
&  =\frac{\alpha}{2a^{2}}\left[  1+\omega\left(  \frac{a_{\text{p}}^{2}}%
{a^{2}}\right)  \right]  +\frac{E_{\text{m}}^{\text{p}}}{a^{3}}. \label{q69}%
\end{align}
so that%

\begin{equation}
\rho-\frac{E_{\text{m}}^{\text{p}}}{a^{3}}=\frac{\alpha}{2a^{2}}\left[
1+\omega\left(  \frac{a_{\text{p}}^{2}}{a^{2}}\right)  \right]  . \label{q70}%
\end{equation}

Accordingly%

\begin{equation}
\rho_{\text{r}}=\frac{\alpha}{2a^{2}}\left[  1+\omega\left(  \frac
{a_{\text{p}}^{2}}{a^{2}}\right)  \right]  ,\text{ \ \ \ \ \ \ \ \ \ \ \ }%
a\geq a_{2}, \label{q71}%
\end{equation}
where $\omega$ is a dimensionless constant and $a_{\text{p}}$ is the present
value of the scale factor. Note that although Eq. (\ref{q31}) for
$\rho_{\text{r}}$ is the same as Eq. (\ref{q15}), its solution (\ref{q71}) is
not the analytic continuation of $\rho$ in Eq. (\ref{q40})$.$ The reason is
that the system is subject to different equations of state in the two regions:
$p=\frac{1}{3}\rho$ for $a_{0}\leq a\leq a_{1}$ and $p=$ $\frac{1}{3}$
$\left(  \rho-E_{\text{m}}^{\text{p}}/a^{3}\right)  $ for $a\geq a_{2}.$These
two regions do not, therefore, belong to the same phase. The region $a_{1}\leq
a\leq a_{2}$ during which rest-mass is created may, therefore, be thought of
as a region of phase transition. This will be further discussed in the next section.

When $\omega$ is expressed in terms of present values of $\rho_{\text{r}}$ and
$a$ one obtains%

\begin{equation}
\omega+1=2\alpha^{-1}a_{\text{p}}^{2}\rho_{\text{r}}^{\text{p}}=\rho
_{\text{r}}^{\text{p}}/\rho_{\text{v}}^{\text{p}}. \label{q72}%
\end{equation}

The total radiation energy $E_{\text{r}}=\rho_{\text{r}}a^{3}$ is then%

\begin{equation}
E_{\text{r}}\thickapprox\frac{\alpha}{2}a+\frac{\alpha\omega}{2}%
\frac{a_{\text{p}}^{2}}{a},\text{ \ \ \ \ \ \ \ \ }a\geq a_{2}. \label{q73}%
\end{equation}

For matter dominance to occur there must exist a value of $a,$ $a=a_{\text{eq}%
}$ say, such that%

\begin{equation}
E_{\text{r}}\left(  a_{\text{eq}}\right)  =E_{\text{m}}^{\text{p}},
\label{q74}%
\end{equation}
where for all $a\geq a_{2},$%

\begin{equation}
E_{\text{r}}\left(  a\right)  \gtrless E_{\text{m}}^{\text{p}}\text{
\ \ \ \ \ when \ }a\lessgtr a_{\text{eq}}. \label{q75}%
\end{equation}
This requires that for $a_{2}\leq a\leq a_{\text{eq}}$ and at least in the
neighborhood of $a_{\text{eq}},$ $a$ must be small enough for $E_{\text{r}}$
to be decreasing at $a=a_{\text{eq}}$ so that%

\begin{equation}
\omega\geq\left(  \frac{a_{\text{eq}}}{a_{\text{p}}}\right)  ^{2}. \label{q76}%
\end{equation}
The value of $a_{\text{eq}}$ is given by%

\begin{equation}
a_{\text{eq}}=\frac{\rho_{\text{m}}^{\text{p}}a_{\text{p}}^{3}\pm\left[
\left(  \rho_{\text{m}}^{\text{p}}a_{\text{p}}^{3}\right)  ^{2}-4\alpha\left(
2\omega\alpha a_{\text{p}}^{2}\right)  \right]  ^{1/2}}{2\alpha} \label{q77}%
\end{equation}
then from Eq. (\ref{q72})%

\begin{equation}
a_{\text{p}}=\frac{\alpha^{1/2}\left(  1+\omega\right)  ^{1/2}}{\sqrt
{2}\left(  \rho_{\text{r}}^{\text{p}}\right)  ^{1/2}} \label{q78}%
\end{equation}

\begin{align}
a_{\text{eq}}  &  =\frac{\rho_{\text{m}}^{\text{p}}\left(  \alpha^{3/2}\left(
1+\omega\right)  ^{3/2}/\sqrt{8}\left(  \rho_{\text{r}}^{\text{p}}\right)
^{3/2}\right)  }{2\alpha}\pm\nonumber\\
&  \frac{\left[  \left(  \rho_{\text{m}}^{\text{p}}\right)  ^{2}\alpha
^{3}\left(  1+\omega\right)  ^{3}/8\left(  \rho_{\text{r}}^{\text{p}}\right)
^{3}-8\alpha^{2}\omega\alpha\left(  1+\omega\right)  /2\rho_{\text{r}%
}^{\text{p}}\right]  ^{1/2}}{2\alpha}\nonumber\\
&  =\frac{1}{\sqrt{8}}\frac{\rho_{\text{m}}^{\text{p}}\alpha^{1/2}\left(
1+\omega\right)  ^{3/2}}{2\left(  \rho_{\text{r}}^{\text{p}}\right)  ^{3/2}%
}\left\{  1\pm\left[  1-\left(  8\right)  \frac{4\omega}{\left(
1+\omega\right)  ^{2}}\left(  \frac{\rho_{\text{r}}^{\text{p}}}{\rho
_{\text{m}}^{\text{p}}}\right)  ^{2}\right]  ^{1/2}\right\}  , \label{q79}%
\end{align}%
\begin{equation}
a_{\text{eq}}^{\text{I}}=\frac{1}{\sqrt{8}}\frac{\rho_{\text{m}}^{\text{p}%
}\alpha^{1/2}\left(  1+\omega\right)  ^{3/2}}{2\left(  \rho_{\text{r}%
}^{\text{p}}\right)  ^{3/2}}\left\{  1-\left[  1-\frac{32\omega}{\left(
1+\omega\right)  ^{2}}\left(  \frac{\rho_{\text{r}}^{\text{p}}}{\rho
_{\text{m}}^{\text{p}}}\right)  ^{2}\right]  ^{1/2}\right\}  . \label{q80}%
\end{equation}

The temperature $T_{\text{eq}}$ at $a=a_{\text{eq}}$ is given by%

\begin{equation}
\rho_{\text{r}}^{\text{p}}=\frac{1}{30}\pi^{2}N\left(  T_{\text{P}}\right)
T_{\text{P}}^{4} \label{q82}%
\end{equation}

\begin{equation}
E_{\text{r}}^{\text{p}}=\rho_{\text{r}}^{\text{p}}a_{\text{p}}^{3}=\frac
{1}{30}\pi^{2}N\left(  T_{\text{P}}\right)  T_{\text{P}}^{4}a_{\text{p}}^{3}
\label{q83}%
\end{equation}

\begin{equation}
E_{\text{r}}\left(  a_{\text{eq}}\right)  =\frac{1}{30}\pi^{2}N\left(
T_{\text{eq}}\right)  T_{\text{eq}}^{4}a_{\text{eq}}^{3} \label{q84}%
\end{equation}

\ Remembering that $E_{\text{r}}\left(  a_{\text{eq}}\right)  =E_{\text{m}%
}^{\text{p}}=\rho_{\text{m}}^{\text{p}}a_{\text{p}}^{3},$ therefore Eq.
(\ref{q84}) becomes%

\begin{equation}
E_{\text{m}}^{\text{p}}=\rho_{\text{m}}^{\text{p}}a_{\text{p}}^{3}=\frac
{1}{30}\pi^{2}N\left(  T_{\text{eq}}\right)  T_{\text{eq}}^{4}a_{\text{eq}%
}^{3}, \label{q85}%
\end{equation}
and%

\begin{equation}
\frac{\rho_{\text{m}}^{\text{p}}}{\rho_{\text{r}}^{\text{p}}}=\frac
{T_{\text{eq}}^{4}a_{\text{eq}}^{3}}{T_{\text{P}}^{4}a_{\text{p}}^{3}},
\label{q86}%
\end{equation}
so that%

\begin{equation}
T_{\text{eq}}^{4}=\left(  \frac{\rho_{\text{m}}^{\text{p}}}{\rho_{\text{r}%
}^{\text{p}}}\right)  \left(  \frac{a_{\text{p}}}{a_{\text{eq}}}\right)
^{3}T_{\text{P}}^{4} \label{q87}%
\end{equation}
with%

\begin{equation}
a_{\text{p}}=\frac{\alpha^{1/2}\left(  1+\omega\right)  ^{1/2}}{\sqrt
{2}\left(  \rho_{\text{r}}^{\text{p}}\right)  ^{1/2}}. \label{q88}%
\end{equation}

Now from Eq. (\ref{q79})%

\begin{equation}
\frac{a_{\text{p}}}{a_{\text{eq}}}=\left[  \frac{\alpha\left(  1+\omega
\right)  }{2\rho_{\text{r}}^{\text{p}}}\right]  ^{1/2}\left[  \frac
{\rho_{\text{m}}^{\text{p}}\alpha^{1/2}\left(  1+\omega\right)  ^{3/2}}%
{2\sqrt{8}\left(  \rho_{\text{r}}^{\text{p}}\right)  ^{3/2}}\left\{
1\pm\left[  1-\frac{\left(  8\right)  \times4\omega}{\left(  1+\omega\right)
^{2}}\left(  \frac{\rho_{\text{r}}^{\text{p}}}{\rho_{\text{m}}^{\text{p}}%
}\right)  ^{2}\right]  ^{1/2}\right\}  \right]  \label{q89}%
\end{equation}

\bigskip%

\begin{equation}
\frac{a_{\text{p}}}{a_{\text{eq}}}=\frac{1}{1+\omega}\left(  \frac
{\rho_{\text{r}}^{\text{p}}}{\rho_{\text{m}}^{\text{p}}}\right)  \left\{
1\pm\left[  1-\left(  8\right)  \frac{4\omega}{\left(  1+\omega\right)  ^{2}%
}\left(  \frac{\rho_{\text{r}}^{\text{p}}}{\rho_{\text{m}}^{\text{p}}}\right)
^{2}\right]  ^{1/2}\right\}  . \label{q90}%
\end{equation}
substituting this into (\ref{q87}), we have%

\begin{equation}
T_{\text{eq}}=\left(  \frac{\rho_{\text{m}}^{\text{p}}}{\rho_{\text{r}%
}^{\text{p}}}\right)  ^{1/4}\left(  \frac{a_{\text{p}}}{a_{\text{eq}}}\right)
^{3/4}T_{\text{P}}. \label{q91}%
\end{equation}
so that%

\begin{equation}
T_{\text{eq}}=T_{\text{P}}\left(  \frac{\rho_{\text{r}}^{\text{p}}}%
{\rho_{\text{m}}^{\text{p}}}\right)  ^{1/2}\left(  1+\omega\right)
^{-3/4}\left\{  1\pm\left[  1-\frac{32\omega}{\left(  1+\omega\right)  ^{2}%
}\left(  \frac{\rho_{\text{r}}^{\text{p}}}{\rho_{\text{m}}^{\text{p}}}\right)
^{2}\right]  ^{1/2}\right\}  ^{-3/4}. \label{q92}%
\end{equation}
\bigskip The radiation energy density $\rho_{\text{r}}^{\text{p}}$ today is
given by%

\begin{equation}
\rho_{\text{r}}^{\text{p}}=\frac{1}{30}\pi^{2}N\left(  T_{\text{P}}\right)
T_{\text{P}}^{4}, \label{q94}%
\end{equation}
where as shown in the Appendix A, $\mathbf{N}\left(  T_{\text{P}}\right)
\mathbf{=}\frac{43}{11}$ in the present model (for neutrino filled universe).
The observational value of $T_{\text{P}}=2.7%
{{}^\circ}%
$ \textbf{K} yields%

\begin{equation}
\rho_{\text{r}}^{\text{p}}=3.8\times10^{-51}\text{ }\left(  \text{GeV}\right)
^{4}. \label{q95}%
\end{equation}

The total energy density today is%

\begin{equation}
\rho^{\text{p}}=\alpha H_{\text{P}}^{2}\thickapprox4\times10^{-47}\text{
}\left(  \text{GeV}\right)  ^{4}, \label{q96}%
\end{equation}
where we have used $H_{\text{P}}=72$ km s$^{-1}$ Mpc$^{-1}.$ This shows that
the universe is matter-dominated at present with $\rho_{\text{m}}^{\text{p}%
}\thickapprox\rho^{\text{p}}.$ This feature is not permanent in the model
under consideration, since a second era of radiation-dominance starts when%

\begin{equation}
a=a_{\text{eq}}^{\text{II}}=\frac{1}{\sqrt{8}}\frac{\rho_{\text{m}}^{\text{p}%
}\alpha^{1/2}\left(  1+\omega\right)  ^{3/2}}{2\left(  \rho_{\text{r}%
}^{\text{p}}\right)  ^{3/2}}\left\{  1+\left[  1-\frac{32\omega}{\left(
1+\omega\right)  ^{2}}\left(  \frac{\rho_{\text{r}}^{\text{p}}}{\rho
_{\text{m}}^{\text{p}}}\right)  ^{2}\right]  ^{1/2}\right\}  . \label{q97}%
\end{equation}

With the ratio $\rho_{\text{r}}^{\text{p}}/\rho_{\text{m}}^{\text{p}%
}\thickapprox1\times10^{-4}$ one can approximate the expressions for
$a_{\text{eq}},$ $T_{\text{eq}}$ and $a_{\text{p}},$ obtaining%

\begin{equation}
a_{\text{eq}}\thickapprox\sqrt{8}\frac{\omega}{\left(  1+\omega\right)
^{1/2}}\frac{\alpha^{1/2}\left(  \rho_{\text{r}}^{\text{p}}\right)  ^{1/2}%
}{\rho_{\text{m}}^{\text{p}}}\thickapprox\frac{\omega}{\left(  1+\omega
\right)  ^{1/2}}\times\frac{\sqrt{8}}{2}, \label{q98}%
\end{equation}
and%
\[
T_{\text{eq}}\thickapprox\frac{1}{\sqrt{8}}\left(  \frac{1+\omega}{\omega
}\right)  ^{3/4}\left(  \frac{\rho_{\text{m}}^{\text{p}}}{\rho_{\text{r}%
}^{\text{p}}}\right)  T_{\text{P}}\thickapprox\frac{2}{\sqrt{8}}\left(
\frac{1+\omega}{\omega}\right)  ,
\]
and%

\begin{equation}
a_{\text{p}}\thickapprox\left(  1+\omega\right)  ^{1/2}\frac{\alpha^{1/2}%
}{\sqrt{2}\left(  \rho_{\text{r}}^{\text{p}}\right)  ^{1/2}}\thickapprox
\left(  1+\omega\right)  ^{1/2}\times\frac{0.7}{\sqrt{2}}\times10^{44}\text{
}\left(  \text{GeV}\right)  ^{-1}. \label{q100}%
\end{equation}
From these we deduce that the $1/a^{4}$ term in $\rho_{\text{r}}$ is dominant
at $a=a_{\text{eq}}$, since%

\begin{equation}
\frac{\omega a_{\text{p}}^{2}}{a_{\text{eq}}^{2}}\thickapprox\frac{\left(
1+\omega\right)  ^{2}}{\omega}\times\frac{1}{4}\times3\times10^{7},
\label{q101}%
\end{equation}
for all $\omega.$ This implies that one has $1/a^{4}$ dominance throughout
$a_{1}\leq a\leq a_{\text{eq}}$ except perhaps for the matter generation
periods. Dominance of $\rho_{\text{r}}$ by this term extends well beyond
$a_{\text{eq}}$ to values of $a$ such that $a\thickapprox10^{-1}\omega
^{1/2}a_{\text{p}}\gg a_{\text{eq}}.$ However, to decide which of the two
terms in $\rho_{\text{r}},$ if any, is dominant at present would depend upon
the value of the parameter $\omega.$

One notes that with $1/a^{4}$ dominance of $\rho_{\text{r}}$ one approximately
has $aT$\textbf{\ }constant as in the standard model. However this does not
imply that the entropy is constant under these conditions, since the variation
of $aT$ \ over the whole\textbf{\ }range\textbf{\ }amounts to considerable
generation of entropy. In the present model there does not exist an entropy
problem in any case since the entropy is initially zero. In the pure radiation
era one has%

\begin{equation}
\frac{dS}{da}=\frac{\alpha}{T}, \label{q102}%
\end{equation}
with%

\begin{equation}
\rho=\frac{1}{30}\pi^{2}NT^{4}=\frac{\alpha}{2a^{2}}\left(  a^{2}-a_{0}%
^{2}\right)  , \label{q1103}%
\end{equation}
and%

\begin{equation}
T=\left(  \frac{15\alpha}{\pi^{2}N}\right)  ^{1/4}\left(  \frac{1}{a}\right)
\left(  a^{2}-a_{0}^{2}\right)  ^{1/4}, \label{q104}%
\end{equation}
we obtain%

\begin{equation}
\frac{dS}{da}=\left(  \frac{\pi^{2}N}{15\alpha}\right)  ^{1/4}\frac{\alpha
a}{\left(  a^{2}-a_{0}^{2}\right)  ^{1/4}}. \label{q105}%
\end{equation}
so that%

\begin{equation}
S=\left(  \frac{1}{8}\right)  ^{3/4}\left[  \frac{4}{3}\left(  \frac{\pi
^{2}N\alpha^{3}}{30}\right)  ^{1/4}\left(  a^{2}-a_{0}^{2}\right)
^{3/4}\right]  , \label{q107}%
\end{equation}
where for $a\geq a_{2},$ i.e. after matter generation, the entropy will be%

\begin{equation}
S=\left(  \frac{1}{8}\right)  ^{3/4}\left[  \frac{4}{3}\left(  \frac{\pi
^{2}N\alpha^{3}}{30}\right)  ^{1/4}\left(  a^{2}+\omega a_{\text{p}}%
^{2}\right)  ^{3/4}\right]  +\text{constant.} \label{q108}%
\end{equation}

This gives for the total entropy generated during the era $a\geq a_{2}$ the expression%

\[
S\left(  a_{\text{p}}\right)  -S\left(  a_{\text{2}}\right)  \thickapprox
\left(  \frac{1}{8}\right)  ^{3/4}\left[  \frac{4}{3}\left(  \frac{\pi
^{2}N\alpha^{3}}{30}\right)  ^{1/4}\left(  \left(  1+\omega\right)
^{3/4}-\left(  \omega\right)  ^{3/4}\right)  ^{3/4}\right]  ,
\]
which means that%

\begin{equation}
S\left(  a_{\text{p}}\right)  -S\left(  a_{\text{2}}\right)  \thickapprox
\left(  \frac{1}{8}\right)  ^{3/4}\times2\left(  1+\omega\right)
^{3/4}\left(  \left(  1+\omega\right)  ^{3/4}-\left(  \omega\right)
^{3/4}\right)  \times10^{93} \label{q109}%
\end{equation}

It is thus seen that a lot of entropy is produced since the end of the matter
generation era. We remark that although the parameter $\omega$ occurs in all
expressions for \textbf{\ }$a_{\text{eq}},$ $T_{\text{eq}},$ $a_{\text{p}}$
and \textbf{\ }$S\left(  a_{\text{p}}\right)  ,$ it is possible to eliminate
it and obtain relations between these quantities. One may also obtain bounds
on these quantities that hold for all $\omega,$ such as for example,%

\begin{equation}
a_{\text{p}}\geq5\times10^{43}\left(  \text{GeV}\right)  ^{-1}. \label{q110}%
\end{equation}
This bound leads to an upper bound on the present value of the cosmological constant,%

\begin{equation}
\lambda_{\text{p}}\leq6\times10^{-88}\text{ }\left(  \text{GeV}\right)  ^{2},
\label{q111}%
\end{equation}
which is well within the upper limit of $10^{-82}$ $\left(  \text{GeV}\right)
^{2}$ placed on $\lambda_{\text{p}}$ from cosmological observation \cite{S.W}.

We now consider the time variation of $a$ for $a\geq a_{2}.$ The field
equation (\ref{q14})%

\[
\left(  \frac{\overset{.}{a}}{a}\right)  ^{2}=\alpha^{-1}\left[  \rho^{\text{r
}}+\rho^{\text{m}}\right]  +\alpha^{-1}\Lambda\left(  t\right)  -\frac
{1}{2a^{2}},
\]
or%

\begin{equation}
\left(  \frac{\overset{.}{a}}{a}\right)  ^{2}=\frac{\alpha^{-1}E_{\text{m}%
}^{\text{p}}}{a^{3}}+\frac{\omega}{2a^{2}}\left(  \frac{a_{\text{p}}^{2}%
}{a^{2}}\right)  +\frac{1}{2a^{2}}, \label{q112}%
\end{equation}
may be written in the form%

\[
\frac{da}{dt}=\left(  \frac{a_{\text{p}}}{a}\right)  \left[  a_{\text{p}%
}\left(  \frac{\rho_{\text{r}}^{\text{p}}}{\alpha}\right)  ^{1/2}\right]
\left[  \frac{1}{2\left(  1+\omega\right)  }\left(  \frac{a}{a_{\text{p}}%
}\right)  ^{2}+\left(  \frac{a}{a_{\text{p}}}\right)  \left(  \frac
{\rho_{\text{m}}^{\text{p}}}{\rho_{\text{r}}^{\text{p}}}\right)  +\frac
{\omega}{2\left(  1+\omega\right)  }\right]  .
\]

Let $\left(  a_{\text{p}}/a\right)  =x$ and $dx=d\left(  a_{\text{p}%
}/a\right)  , $ therefore%

\[
\int_{0}^{t}dt%
\acute{}%
=\left(  \frac{\alpha}{\rho_{\text{r}}^{\text{p}}}\right)  ^{1/2}\int
_{0}^{a_{\text{p}}/a}\frac{xdx}{\left[  2^{-1}\left(  1+\omega\right)
^{-1}x^{2}+\left(  \rho_{\text{m}}^{\text{p}}/\rho_{\text{r}}^{\text{p}%
}\right)  x+\omega/\left[  2\left(  1+\omega\right)  \right]  \right]  ^{1/2}}%
\]

\bigskip%

\begin{equation}
t=\left(  \frac{\alpha}{\rho_{\text{r}}^{\text{p}}}\right)  ^{1/2}\int
_{0}^{a_{\text{p}}/a}\frac{xdx}{\left[  \frac{1}{2\left(  1+\omega\right)
}x^{2}+\left(  \rho_{\text{m}}^{\text{p}}/\rho_{\text{r}}^{\text{p}}\right)
x+\frac{\omega}{2\left(  1+\omega\right)  }\right]  ^{1/2}} \label{q113}%
\end{equation}
under the approximate boundary condition $t=t_{2}\thickapprox0,$ when
$a=a_{2}\thickapprox0.$ For the present age $t_{\text{p}}$ of the
matter-dominated universe one obtains, for all $\omega,$%

\begin{equation}
\left(  \frac{\alpha}{\rho_{\text{r}}^{\text{p}}}\right)  ^{1/2}\int_{0}%
^{1}\frac{xdx}{\left[  \left(  \rho_{\text{m}}^{\text{p}}/\rho_{\text{r}%
}^{\text{p}}\right)  x+1/2\right]  ^{1/2}}\leq t_{\text{p}}\left(
\omega\right)  \leq\left(  \frac{\alpha}{\rho_{\text{r}}^{\text{p}}}\right)
^{1/2}\int_{0}^{1}\frac{x^{1/2}dx}{\left[  \left(  \rho_{\text{m}}^{\text{p}%
}/\rho_{\text{r}}^{\text{p}}\right)  +\left(  1/2\right)  x\right]  ^{1/2}},
\label{q114}%
\end{equation}

Another parameter which is almost $\omega$- independent is the present value
$q_{\text{p}}$ of the deceleration parameter defined by $q=-a\overset{..}%
{a}/\overset{.}{a}^{2}.$ One obtains for this value%

\begin{equation}
q=-\frac{a\overset{..}{a}}{\overset{.}{a}^{2}}=-\left(  \frac{\overset{..}{a}%
}{a}\right)  \left(  \frac{a}{\overset{.}{a}}\right)  ^{2}. \label{q115}%
\end{equation}
Substituting for $\left(  \frac{\overset{..}{a}}{a}\right)  $ and $\left(
\frac{a}{\overset{.}{a}}\right)  ^{2}$ from the above we obtain%

\begin{equation}
q_{\text{p}}=\frac{1}{2}+\left(  \frac{\omega}{\left(  1+\omega\right)
}-\frac{1}{2}\right)  \left(  \frac{\rho_{\text{r}}^{\text{p}}}{\rho
_{\text{m}}^{\text{p}}}\right)  +......, \label{q116}%
\end{equation}
so that $q_{\text{p}}\thickapprox\frac{1}{2}$ in this model. This is nearly
the same value as in the standard model with $k=0.$

We now consider the period of generation of rest-mass and in particular show
that in this model the pressure must have been negative during part of this period.

\subsection{Period of matter generation}

Consider Eq. (\ref{q15}) with $\Lambda$ given by (\ref{q36})%

\[
dE+pda^{3}=\alpha da,
\]
where we have written $E$ for the total energy $\rho a^{3}.$ This equation is
valid for all $a.$ Integrating it between $a=a_{0}$ and $a=a_{2},$ the end of
the matter generation period, one obtains:%

\begin{equation}
3\int_{a_{0}}^{a_{2}}pa^{2}da=\alpha\left(  a_{2}-a_{0}\right)  -E_{2},
\label{q117}%
\end{equation}
where $E_{2}=E\left(  a_{2}\right)  $ and we used the initial condition
$E\left(  a_{0}\right)  =0.$ Now,%

\begin{equation}
E_{2}=E_{\text{m}}\left(  a_{\text{p}}\right)  +E_{\text{r}}\left(
a_{2}\right)  , \label{q118}%
\end{equation}
since we assume that the rest-mass contribution $E_{\text{m}}$ to the total
energy remains constant for all $a\geq a_{2}.$ Using Eqs. (\ref{q66}),
(\ref{q72}) and (\ref{q73}), Eq. (\ref{q117}) becomes
\begin{align}
E_{2}  &  =E_{\text{m}}\left(  a_{\text{p}}\right)  +E_{\text{r}}\left(
a_{2}\right) \nonumber\\
&  =\rho_{\text{m}}^{\text{p}}a_{\text{p}}^{3}+\rho_{\text{r}}\left(
a_{2}\right)  a_{2}^{3}\nonumber\\
&  =\frac{\alpha a_{2}}{2}\left[  \left(  \frac{\rho_{\text{m}}^{\text{p}}%
}{\rho_{\text{r}}^{\text{p}}}\right)  \left(  \frac{a_{\text{p}}}{a_{2}%
}\right)  +1\right]  +\frac{\alpha\omega}{2}\frac{a_{\text{p}}^{2}}{a_{2}}.
\label{q199}%
\end{align}
Substituting \ Eq. $\left(  \text{\ref{q199}}\right)  $ into $\left(
\text{\ref{q117}}\right)  $ we obtain%

\begin{equation}
3\int_{a_{0}}^{a_{2}}pa^{2}da=\alpha\left(  a_{2}-a_{0}\right)  -E_{2}%
=-\frac{\alpha a_{2}}{2}\left[  \left(  1+\omega\right)  \left(  \frac
{\rho_{\text{m}}^{\text{p}}}{\rho_{\text{r}}^{\text{p}}}\right)  \left(
\frac{a_{\text{p}}}{a_{2}}\right)  -1\right]  -\frac{\alpha\omega}{2}%
\frac{a_{\text{p}}^{2}}{a_{2}}-\alpha a_{0}. \label{q120}%
\end{equation}

The right-hand side of (\ref{q120}) is clearly negative since $\rho_{\text{m}%
}^{\text{p}}\geq\rho_{\text{r}}^{\text{p}}$, thus the integral on the
left-hand side of this equation must be negative. But%

\begin{equation}
\int_{a_{0}}^{a_{2}}pa^{2}da\geq0 \label{q121}%
\end{equation}
indicating that the pressure must have been negative during part of the region
of matter generation.

As mentioned before, the region of matter generation is a phase transition
period between region of known and different equations of state. It now
appears that the equation of state associated with the creation of rest-mass
is characterized by negative pressure. Ideas on phase transitions in the early
universe from unified gauge field theories indicate that phase transitions are
expected to have occurred at least twice; at $T$ $\sim10^{15} $ GeV (GUT phase
transition) and at $T$ $\sim10^{2}$ GeV (electro-weak phase transition)$.$ If
the maximum temperature is of order $10^{19}$ GeV, this gives $a_{1}$
$\lesssim10^{-10}$ $\left(  \text{GeV}\right)  ^{-1}$ so that $t_{1}%
\leq10^{-34}$ s.

We now observe that the period of matter generation separates the pure
radiation regime in which $\overset{..}{a}\geq0$ and matter-and-radiation
regime in which $\overset{..}{a}\leq0$ as may be seen as%

\[
\frac{\ddot{a}}{a}=-\frac{4\pi}{3}\text{ }\left(  \rho_{^{\text{effective}}%
}+3p\right)  +\frac{\lambda_{^{\text{effective}}}}{3}.
\]
This can be written as%

\begin{equation}
\frac{\ddot{a}}{a}=\alpha^{-1}\left[  \frac{2\alpha}{2a^{2}}-\frac{1}{2}\text{
}\left(  \rho+3p\right)  \right]  , \label{q122}%
\end{equation}
which means that%

\begin{equation}
a\ddot{a}=1-\frac{1}{2\alpha}\text{ }\left(  \rho+3p\right)  a^{2},
\label{q123}%
\end{equation}
and that, for $a\geq a_{2}$%

\begin{equation}
\frac{1}{2\alpha}\text{ }\left(  \rho+3p\right)  a^{2}=\frac{1}{2}%
+\frac{\omega}{2}\frac{a_{\text{P}}^{2}}{a^{2}}+\frac{\rho_{\text{m}}}{\alpha
}a^{2}\geq1. \label{q124}%
\end{equation}

It thus appears that the creation of rest-mass results in the reversal of the
sign of \ $\ddot{a}.$ Thus one might say that the presence of rest-mass causes
the deceleration of the expansion of the universe. The rate of generation of
energy is given by the equation%

\begin{equation}
\frac{dE}{da}=\alpha-3pa^{2}. \label{q125}%
\end{equation}
This shows that the maximum rate of energy generation occurs within the
negative pressure interval after the pressure has attained its maximum
negative value.

\section{Discussion}

Motivated by previous studies of the back-reaction effect of quantum fields in
the Einstein static model for the universe we have tried to construct a
working model for a dynamic universe. The background geometry is assumed to be
that of the closed Robertson-Walker model. We assumed the existence of a
non-zero cosmological constant for start and have solved the Einstein field
equations accordingly. Quantum field are shown to produce a non-zero vacuum
energy density which will be a source for a critical density dynamic universe
that will continue expanding without limit. The non-zero value of the vacuum
expectation value of the energy-momentum tensor of the quantum field causes
the universe to have a non-singular start, though the radius is of the size of
Planck length. The critical density universe was originally considered in
Friedmann cosmology to indicate a spatially flat universe, but since we have
introduced a non-zero cosmological constant here then it is legitimate not to
consider the universe to be exactly flat but to be nearly flat. This might
explain the recent analysis of the CMB measurements which showed that the
universe is nearly flat rather than being exactly flat. The subsequent
development of the universe is controlled by the back-reaction effects. Our
model is free from all the standard big bang model problems and is
non-singular. However, it is found that there should be a continuos creation
of matter and energy in this universe at a rate lower than that proposed by
the steady-state theory. The rate of matter/energy creation is proportional to
the radius of the universe and consequently the overall density of universe
drops as $1/a^{2}$. This creation of matter/energy can be explained by the
fact of the conversion of the energy contained in the cosmological constant
into particles and radiation. We did not attempt to suggest a mechanism for
the creation of such energy but we have determined the necessary relations to
the cosmological constant.\bigskip

\textbf{Figure Caption:}

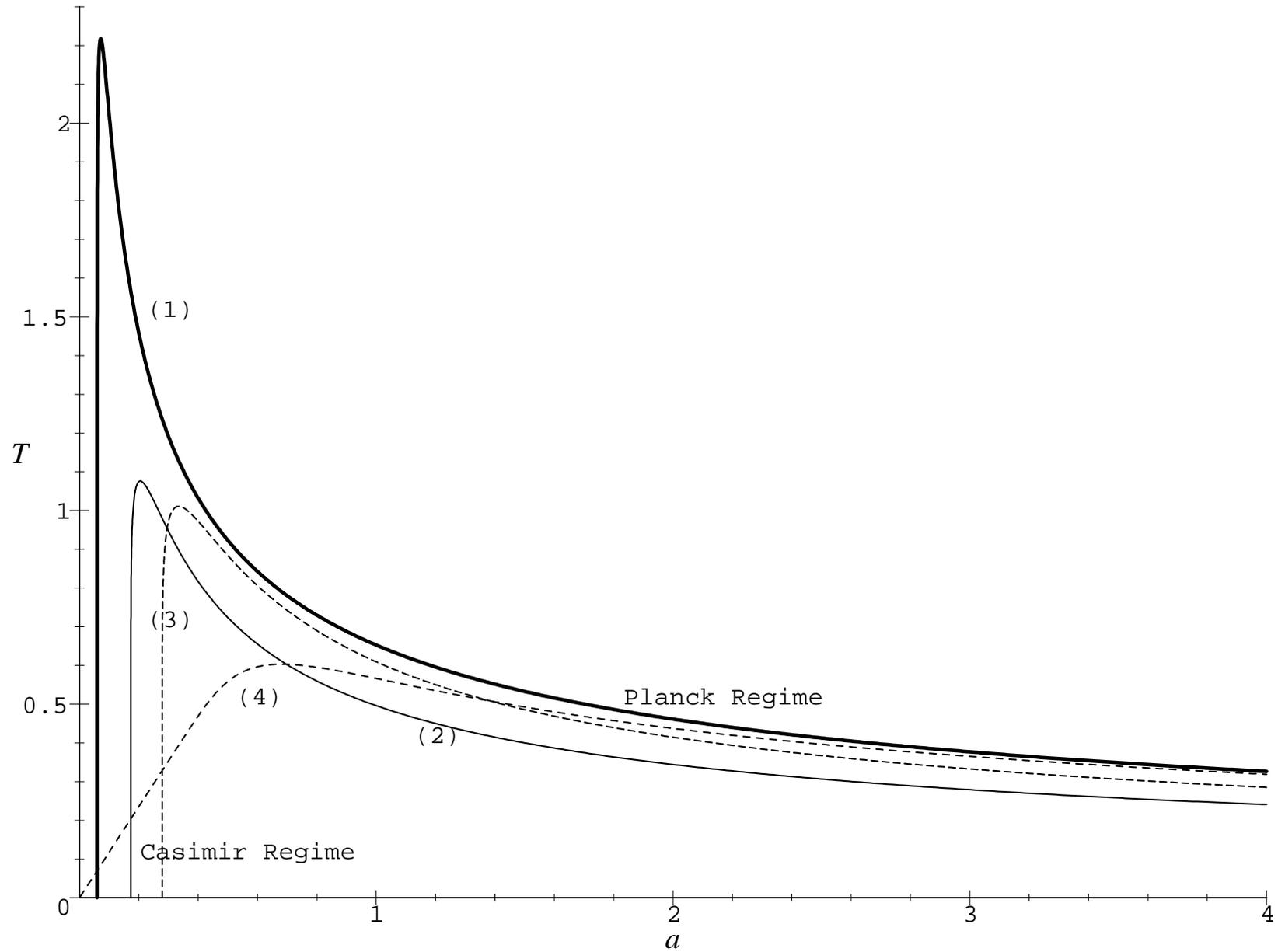
Fig. 1: The temperature-radius relationship deduced from the back-reaction
effect of the vacuum energy ploted for different matter fields: the
conformally coupled scalar field (1), the neutrino field (2), the photon field
(3) and the minimally coupled field (4) (see ref. \cite{Altaie4}). 

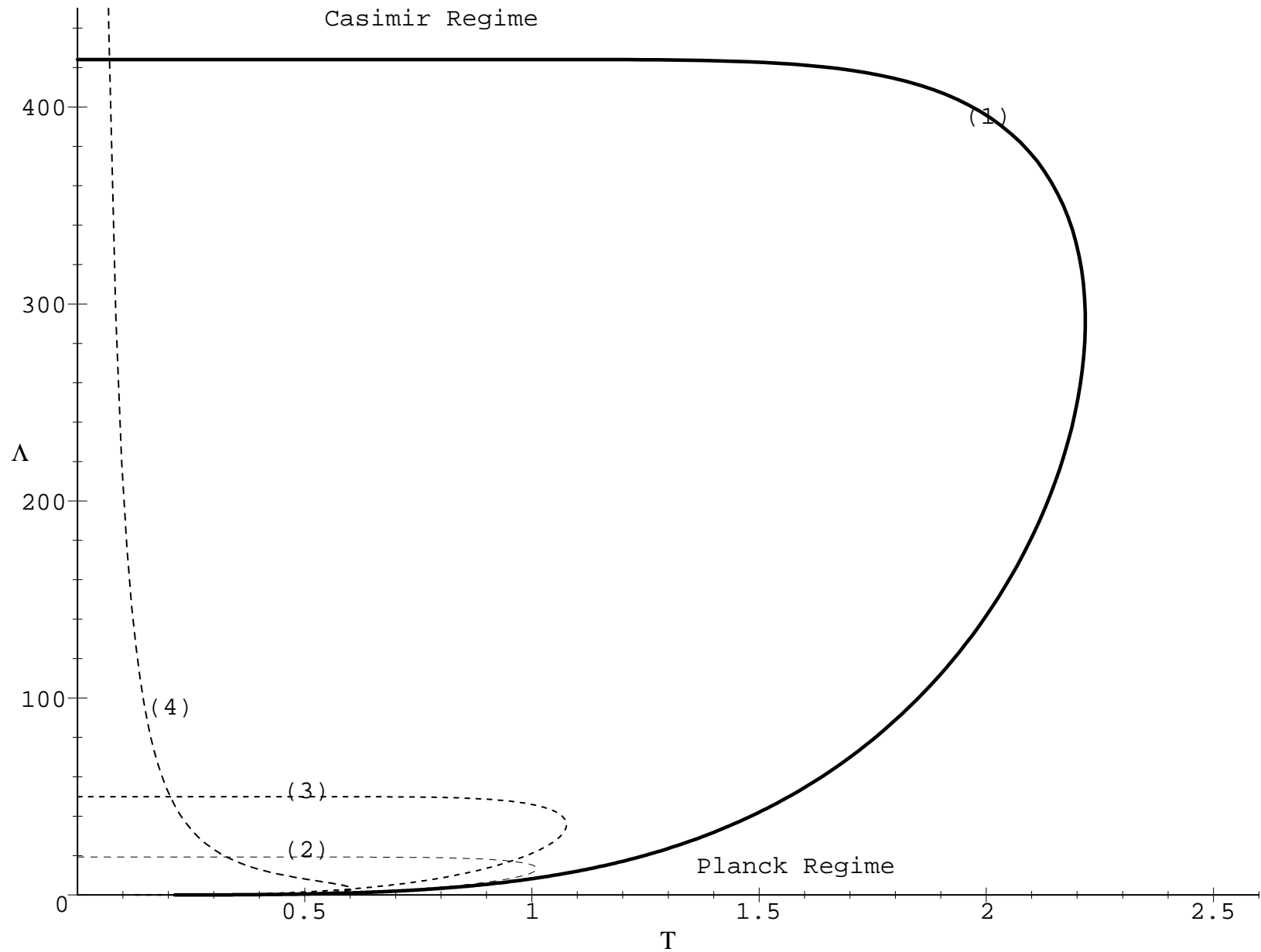
Fig. 2: Depicts the contributions of the conformally coupled scalar field (1),
the photon field (2), the neutrino field (3) and the minimally coupled scalar
field (4) to the cosmological constant in an Einstein universe at finite
temperatures (see ref. \cite{Altaie4}).

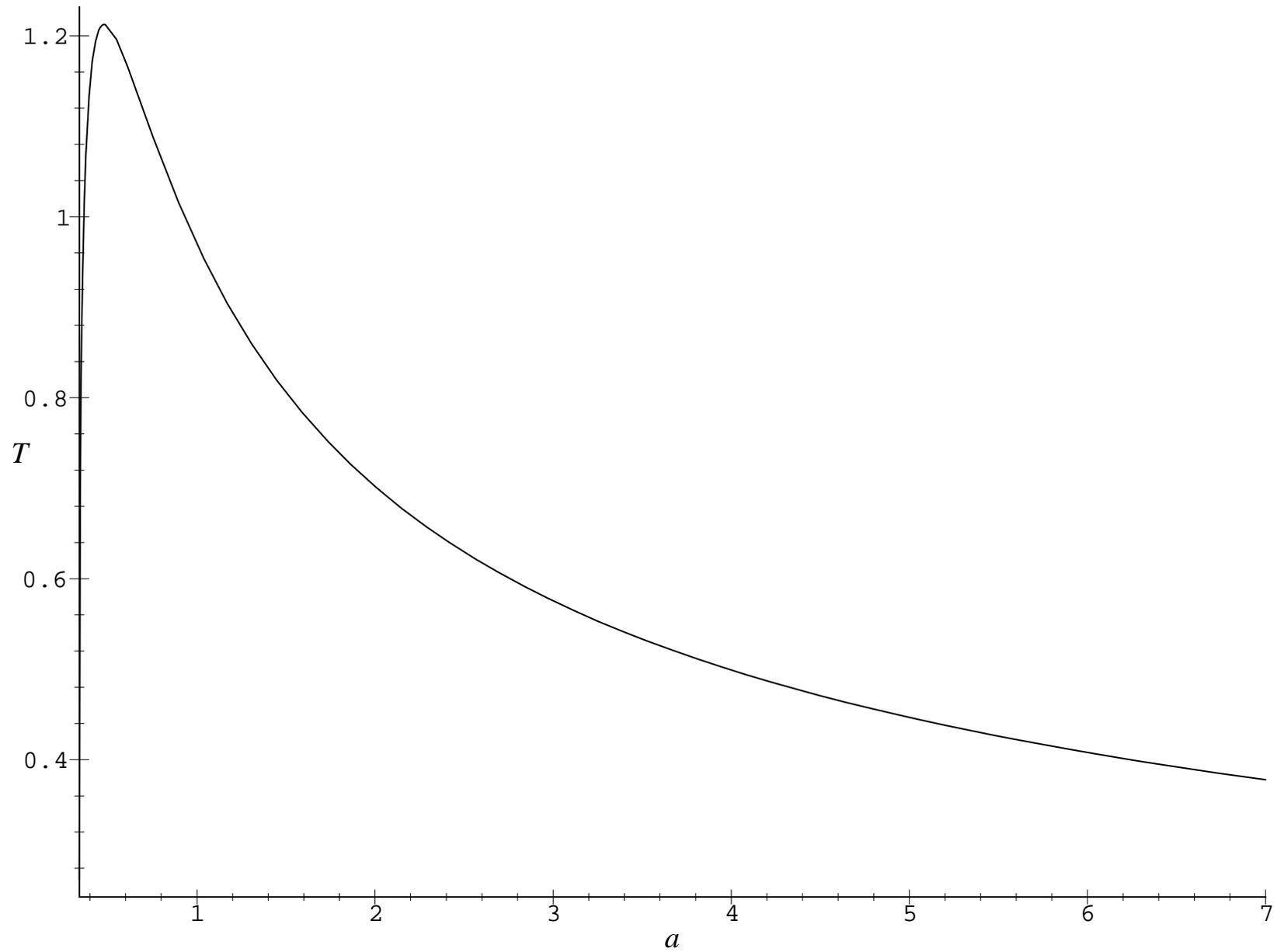
Fig. 3: The temperature-radius relationship according to this model. Note that
the $x$-axis does not start from zero but from $0.34$.

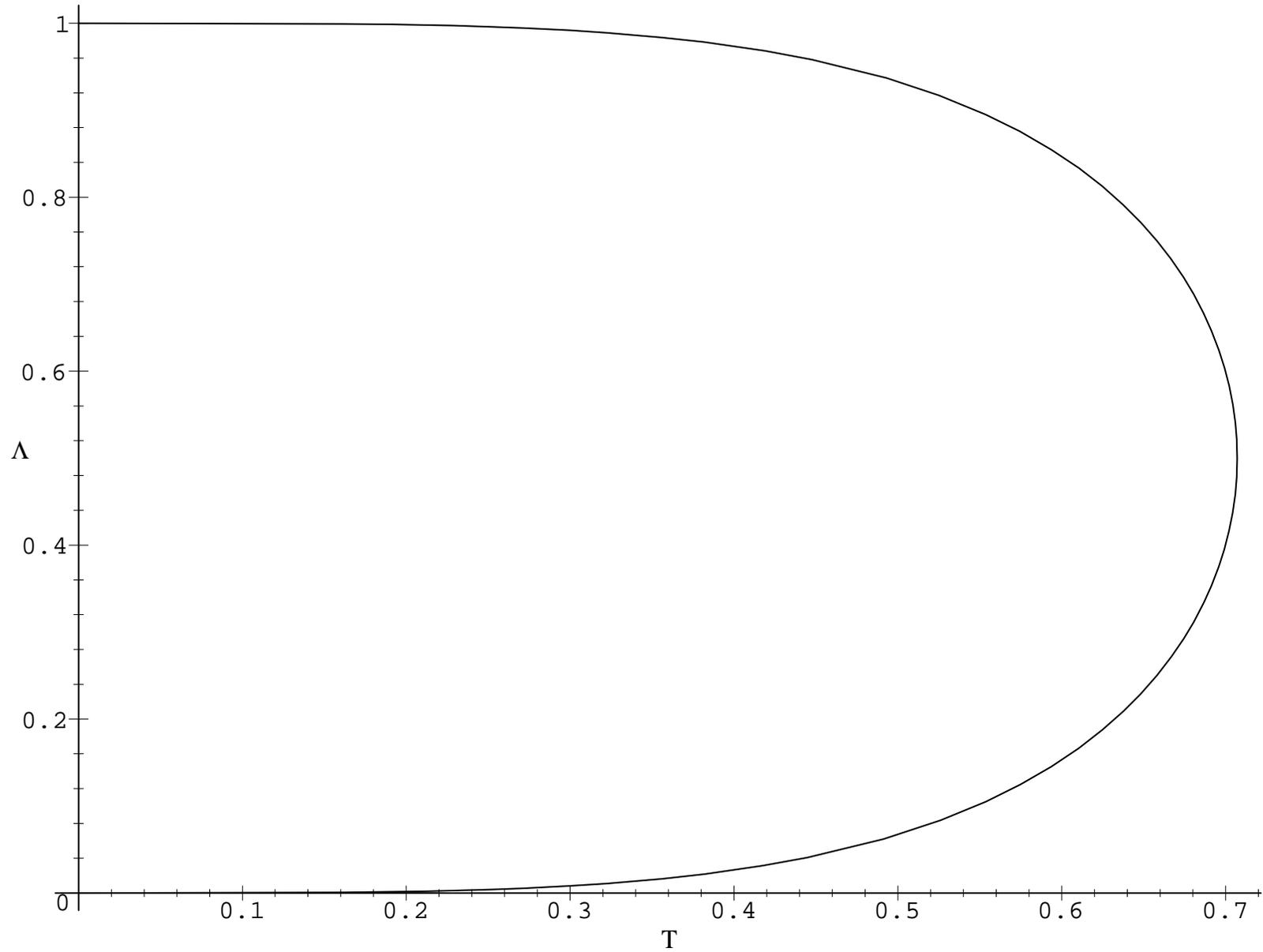
Fig. 4: The temperature dependence of the cosmological constant calculated in
accordance with the present model.

\end{document}